\documentclass[iop]{emulateapj}

\usepackage{graphicx}
\usepackage{amsmath}
\usepackage{subfigure}
\usepackage{float}
\usepackage{natbib} 
\usepackage{hyperref}
\usepackage{tabularx}
\usepackage[T1]{fontenc}

\renewcommand{\dataset}[1]{doi:{#1}, \url{https://doi.org/#1}}

\begin{document}

\title{No Such Thing as a Simple Flare: Substructure and QPPs Observed in a Statistical Sample of 20 Second Cadence TESS Flares}

\author{Ward S. Howard\altaffilmark{1}, Meredith A. MacGregor\altaffilmark{1}}
\altaffiltext{1}{Department of Astrophysical and Planetary Sciences, University of Colorado, 2000 Colorado Avenue, Boulder, CO 80309, USA}

\email[$\star$~E-mail:~]{Ward.Howard@colorado.edu}

\begin{abstract}
A 20 second cadence TESS monitoring campaign of 226 low-mass flare stars during Cycle 3 recorded 3792 stellar flares of $\geq$10$^{32}$ erg. We explore the time-resolved emission and substructure in 440 of the largest flares observed at high signal/noise, 97\% of which released energies of $\geq$10$^{33}$ erg. We discover degeneracy present at 2 minute cadence between sharply-peaked and weakly-peaked flares is common, although 20 second cadence breaks these degeneracies. We better resolve the rise phases and find 46\% of large flares exhibit substructure during the rise phase. We observe 49 candidate quasi-periodic pulsations (QPP) and confirm 17 at $\geq$3$\sigma$. Most of our QPPs have periods less than 10 minutes, suggesting short period optical QPPs are common. We find QPPs in both the rise and decay phases of flares, including a rise-phase QPP in a large flare from Proxima Cen. We confirm the \citet{davenport2014} template provides a good fit to most classical flares observed at high cadence, although 9\% favor Gaussian peaks instead. We characterize the properties of complex flares, finding 17\% of complex flares exhibit ``peak-bump" morphologies composed of a large, highly impulsive peak followed by a second more gradual Gaussian peak. We also estimate the UVC surface fluences of temperate planets at flare peak and find 1/3 of 10$^{34}$ erg flares reach the D90 dose of \textit{D. Radiodurans} in just 20 seconds in the absence of an atmosphere.
\end{abstract}


\maketitle

\section{Introduction}

M-dwarfs make up 75\% of nearby stars \citep{henry2006, Covey2008} and frequently host rocky planets at habitable zone (HZ) distances that could allow for stable liquid water on their surfaces (e.g. \citealt{Dressing2013, DressingCharbonneau2015}). However, the proximity of the HZ to the host star and frequent powerful flares call into question the habitability of M-dwarf planets \citep{tarter2007}. Flares occur when a star's magnetic field re-connects, releasing radiation across the electromagnetic spectrum \citep{Kowalski2013}. Rocky HZ planets around M-dwarfs are often subjected to superflares \citep{Howard2019, Gunter2020}, i.e. flares emitting 10-1000 times the energy of the largest solar flares \citep{Schaefer2000}. The role that these superflares play in determining the habitability of M-dwarf planets is still debated. Frequent superflares can erode the ozone layer of an Earth-like atmosphere and allow lethal amounts of UV surface flux \citep{Tilley2019}, while a low flare rate may not result in enough UV radiation to power pre-biotic chemistry \citep{Ranjan2017, Rimmer2018}. Alternately, superflares may be preferentially directed away from M-dwarf planets \citep{Ilin2021} or even increase the detectability of bio-signatures \citep{OMalley2019_bio, Chen2021}.

Although M-dwarf superflares clearly play a key role in determining the habitability of exoplanets \citep{Loyd2018, Tilley2019}, many of their detailed time-resolved properties (especially substructure) have not been characterized in the optical in large statistical ensembles with high time resolution. Stellar flare emission is well-known to change on timescales of seconds to tens of seconds \citep{Moffett1974, Kowalski2019}, making observing cadences of $\sim$1 minute or slower less suitable for characterizing the time-resolved properties of flares \citep{Yang2018, Kowalski2019, Jackman2021}. To date, the most comprehensive statistical sample of resolved flares was obtained by \citet{Jackman2021} and measured the $t_\mathrm{1/2}$ or FWHM times and energies of 285 flares from 158 M-dwarfs at 13 second cadence with the Next Generation Transit Survey (NGTS; \citealt{Wheatley2018}). Recently, \citet{Gilbert2021} measured the flare energy for 125 large flares from AU Mic at 20 second cadence with Transiting Exoplanet Survey Satellite (TESS; \citealt{Ricker_TESS}) data. The time-resolved properties and substructure of several specific flares were illustrated in both of these surveys but not systematically quantified.

One of the primary metrics describing the time-resolved properties of stellar flares is their impulsivity (e.g. \citealt{Kowalski2013, Kowalski2016, Silverberg2016, Howard2020b}). The impulse of a flare describes the morphology of the flare light curve and is defined as the amplitude of a flare's peak in fractional-flux units over its FWHM duration in minutes \citep{Kowalski2013}. Because highly impulsive superflares deliver the same amount of UV flux much more rapidly than less impulsive ones, it is likely the temporal structure of the flare peak could influence the rates of key reactions in atmospheric chemistry \citep{Loyd2018} or biology \citep{Abrevaya2020}. Ultra-high-cadence NUV/optical observations by \citet{Kowalski2013, Kowalski2016} find that flares can occur with a range of morphologies from highly impulsive and strongly peaked flares to less impulsive and more gradual events. While impulse measurements for superflares are generally lacking at very high cadences, observations of superflares at 2 min cadence are also consistent with a range of impulse values \citep{Howard2019}. In this work, we use ``ultra high cadence" to refer to cadences of $\sim$1 sec, ``high cadence" to refer to cadences of $\sim$10 sec, ``moderate cadence" for $\sim$100 sec, and anything slower as low cadence.

Flares are often composed of multiple impulsive events. Classical flares are best described by a single flare peak while complex flares are best described by a superposition of flare peaks. Larger flares are more likely to have clearly complex light curve morphologies than smaller ones due to sympathetic emission from nearby active regions or a cascade of emission from the same active region \citep{hawley2014, Davenport2016proc}. In a flare cascade scenario, the same active region emits flares of decreasing size until the stored energy is depleted \citep{Davenport2016proc}. On the other hand, sympathetic flaring occurs when one flare triggers another flare from a spatially-adjacent active region. Sympathetic flaring is regularly observed from the Sun \citep{Schrijver2015} and may be responsible for complex stellar flares \citep{Davenport2016proc, Kowalski2019}.

Multiple peaks within flares sometimes exhibit periodic variability. These quasi-periodic pulsations (QPPs) are oscillations in the flare emission and are observed from both the Sun and other stars \citep{Nakariakov2009, Doorsselaere2016}. The mechanism that excites QPPs in flares is currently uncertain, but both quasi-steady (i.e. magnetic dripping) and magnetohydrodynamic (MHD) processes have been proposed. If either of these mechanisms is correct, flare seismology would be possible. Flare seismology probes quantities such as the size, shape, magnetic field strength, and/or Alfv\'en speed of flare loops \citep{Doorsselaere2016}. By providing these difficult-to-measure quantities, QPPs may in turn lead to better estimates of the particle and radiation environments of exoplanets \citep{Doorsselaere2016}. However, much remains unknown about the occurrence of optical pulsation periods. Most QPPs observed in the optical have periods of $\sim$10 minutes or longer (e.g. \citealt{Anfinogentov2013, Balona2015, Pugh2016, Vida2019, Ramsay2021}). It has proven difficult to determine if this is due to astrophysics or observational bias resulting from low observing cadences \citep{Balona2015, Million2021, Monsue2021}.

The TESS mission has now observed tens of thousands of M-dwarfs at 2 min cadence \citep{Stassun2019}, including extensive observations of M-dwarf flare stars \citep{Vida2019, Doyle2019, Gunter2020, Feinstein2020, Medina2020, Howard2020b, Ramsay2021, Gilbert2021, Pope2021, Howard_Law_2021}. However, 2 minute cadence TESS observations do not consistently resolve the substructure of these flares. Beginning in Cycle 3, a new 20 second cadence mode was introduced as part of TESS's extended mission. The new TESS 20 second cadence mode has already shown great promise in resolving short-period asteroseismic oscillations \citep{Huber2021} and stellar flares from AU Mic \citep{Gilbert2021}. An ongoing 20 second cadence flare monitoring program with TESS observed 226 M-dwarf flare stars during Cycle 3. This program is designed to observe a large statistical sample of rare superflares at high time resolution and high signal/noise in order to probe the emission mechanisms and habitability impacts of superflare events. Large numbers of resolved superflares makes possible detailed characterization of the substructure in these events, shedding light on their stellar emission environments \citep{Kowalski2013}.

In Section \ref{TESSphot}, we describe the TESS observations. In Section \ref{tess_flares} we describe flare identification in the TESS light curves and the construction of our statistical sample of flares. In Section \ref{2min_vs_20sec}, we explore the morphology of our sample of 20 second cadence flare sample. In Section \ref{qpp_flares}, we describe quasi-periodicity in the substructure of the flares. In Section \ref{common_morphologies}, we describe common morphologies among complex flares. In Section \ref{hab_implication}, we explore habitability implications of 20 sec cadence flare emission. In Section \ref{conclude}, we summarize and conclude.

\section{TESS observations}\label{TESSphot}
In its initial and extended missions, TESS is searching for transiting exoplanets and astrophysical variability across the entire sky, split into $\sim$26 sectors. TESS observes each sector continuously with four 10.5 cm optical telescopes in a red (600-1000 nm) bandpass for 28 days at 21$\arcsec$ pixel$^{-1}$. Calibrated, 20 second cadence TESS light curves from TESS Cycle 3 GO 3174 of each sector 27-37 flare star and Proxima Cen were downloaded from the Mikulski Archive for Space Telescopes (MAST) \footnote{https://mast.stsci.edu. The specific observations analyzed can be accessed via \dataset{10.17909/t9-st5g-3177}}. We selected Simple Aperture Photometry (SAP) light curves rather than Pre-search Data Conditioning (PDC) ones to avoid altering the impulsive flare structure or rotational modulation.

TESS observed 226 M-dwarf flare stars spread across the Southern sky in the new 20 second cadence mode as part of Cycle 3 GI Program 3174. The flaring properties of these M-dwarfs have been extensively studied at 2 min cadence as part of the EvryFlare Survey (e.g. \citealt{Howard2018,Howard2019,Howard2020a,Howard2020b, Howard_Law_2021}). The sample is composed of very nearby (23$\substack{+27 \\ -12}$ pc) and bright ($g^{\prime}$=12.3$\substack{+1.4 \\ -1.0}$) flare stars, where the quoted spreads are 1$\sigma$ standard deviations. The nearest system is Proxima Cen at 1.3 pc. These sources are also TESS planet-search targets suitable for future characterization efforts (e.g. \citealt{Wakeford2019, Teske2020}). Spectral types are updated from those of \citet{Howard2020a} with SIMBAD \citep{Wenger2000} classifications where available. Flare stars were selected to cross the fully-convective boundary with 146 M-dwarfs of types $\approx$M0-M3, 47 $\sim$M4 dwarfs, and 16 M5-M6 dwarfs.

\section{Constructing a 20 second cadence sample of large flares}\label{tess_flares}

\begin{figure*}
	\centering
	{
		\includegraphics[trim= 0 0 0 0, width=0.98\textwidth]{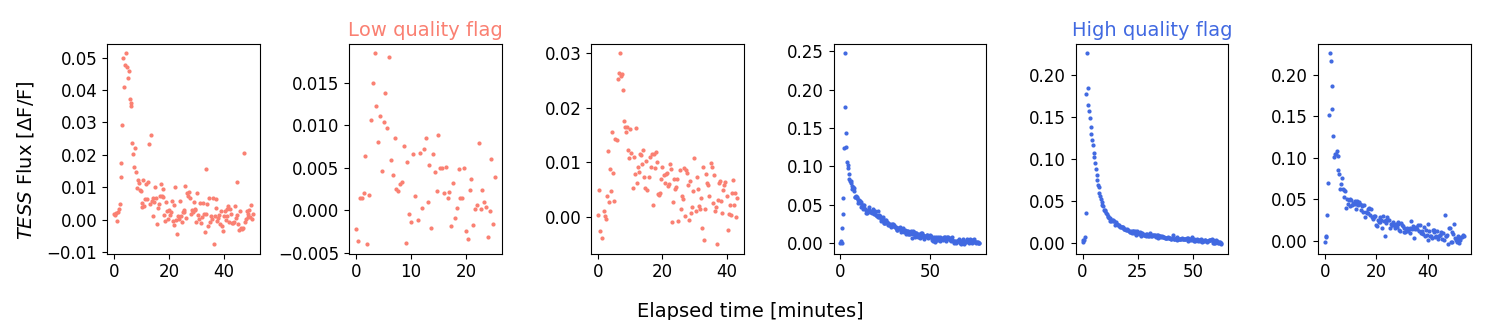}
	}
	\caption{Examples of flares that were flagged as high or low quality during visual inspection. Low quality events are defined as flares where it is difficult to distinguish if substructure is present or not due to high scatter or a brief duration.}
	\label{fig:highlowquality}
\end{figure*}

\subsection{Flare identification and false-positive vetting}\label{flare_identification}
We remove long-term astrophysical and systematic variability from the SAP light curve of each target before searching for flares. We first mask out times corresponding to incorrect background subtractions and other large-scale systematic changes identified in an initial inspection of the light curves. Next, we apply a Savitzky–Golay (SG; \citealt{Savitzky1964}) filter to the light curve in a two-step process to fit and remove astrophysical variability as described in \citet{Howard_Law_2021}. To prevent the SG filter from also fitting flares during the first step, all epochs in close proximity to 2.5$\sigma$ photometric outliers are not used in the fit. During the second step, only epochs within the times of flare candidates reaching $\geq$3$\sigma$ above the photometric noise are excluded from the fit. This two-step process gives remarkable agreement to the out-of-flare variability of our stars. Spot checks were performed across the sample to ensure the model accurately traced the out-of-flare light curves but did not trace the flares. As in \citet{Howard_Law_2021}, when the stellar rotation period $P_\mathrm{rot} <$ 1 d, then the window length is set to 101 points. When 1 $\leq P_\mathrm{rot} <$ 2 d, then it is set to 151 points. If 2 $\leq P_\mathrm{rot} <$ 4 d, then it is set to 201 points. Otherwise, the window length is set to 401 points. The number of points assumes the context of a light curve at 0.3 to 2 min cadence in units of days and are verified by eye to correctly model rotational variability only within this context.

Flares are identified in the light curves by an automated algorithm as $\geq$3$\sigma$ excursions above the photometric scatter of the full light curve excluding flare times after the first de-trending step described above. Visual checks confirmed flares large enough to be clearly visible to the eye were also correctly detected in the de-trended light curves after step one. De-trending step two was necessary for accurate flare amplitudes and energies. The higher cadence led to several modifications of the flare finding algorithm not necessary on 2 min data (e.g. \citealt{Howard2020b}). Because the 20 second cadence light curves exhibit higher scatter than the 2 minute data, the algorithm suppresses large numbers of false positives by only accepting events in which 3 flux measurements occurring within a $\sim$2 minute window exceed 3$\sigma$. The second modification needed for the higher cadence is for determining start and stop times. In \citet{Howard2020b}, start and stop times were assigned as the first and last epochs with fluxes divided by the standard deviation in photometric scatter $\sigma>$0. A $\sim$2 min rolling average of the 20 second cadence data is required to suppress outliers that prevent a determination of the correct noise floor. Flare start and stop times are therefore given at the epochs where the rolling average of the light curve divided by the standard deviation in scatter drops to $\approx$0 before and after the flare peak time. Flare start and stop times are subsequently adjusted by eye if necessary. This was only necessary for 73 out of thousands of candidates. Once the algorithm has identified the candidates, we then vet each candidate by eye, removing the remaining false positives and marginal detections. During visual inspection, we also carefully inspect each flare to ensure out-of-flare variability was correctly subtracted.

\begin{figure*}
	\centering
	{
		\includegraphics[trim= 0 0 0 0, width=0.9\textwidth]{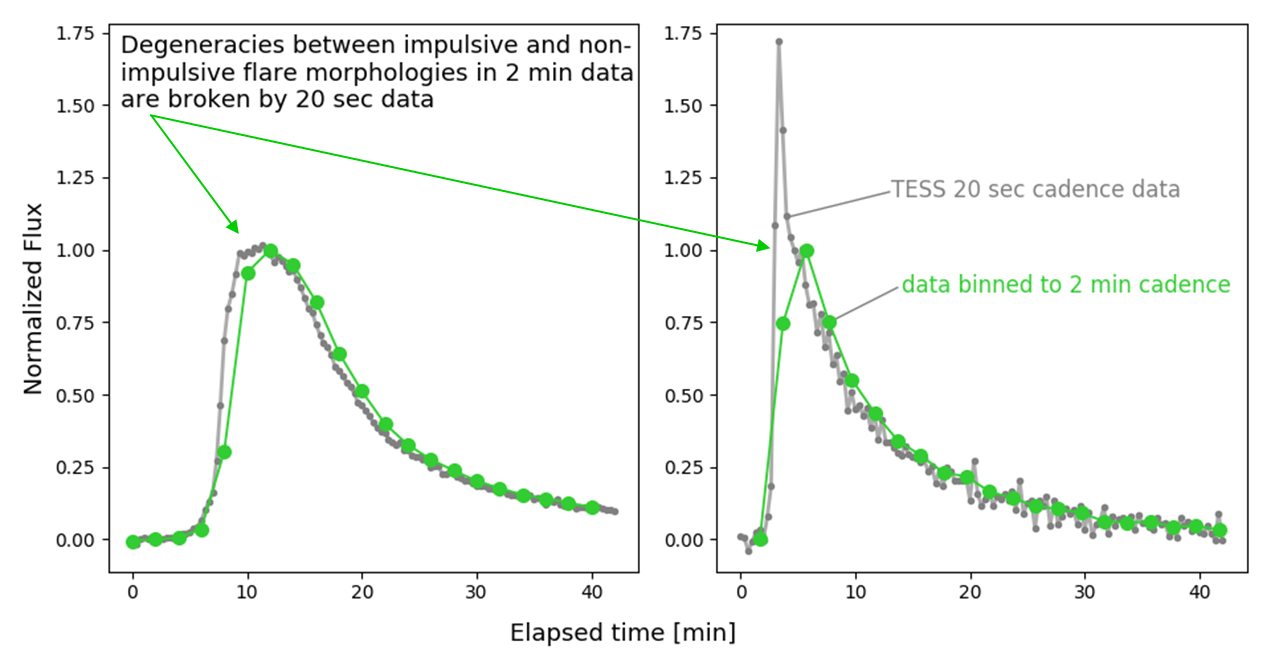}
	}
	\vspace{-0.2cm}
	\caption{Even the morphologies of very large M-dwarf flares can be degenerate at 2 min cadence. While both flares appear similar when binned to 2 min cadence (green), 20 second cadence observations reveal the left flare actually emits at a near-constant level during peak while the right flare emits according to a classical impulsive profile. Since the high cadence behavior of flares is usually assumed to follow a classical emission profile in photo-chemical models of planetary atmospheres and other contexts (e.g. \citealt{Howard2018, Tilley2019}), alternate morphologies may change the rates of chemical reactions as compared to those modelled under classical flares.}
	\label{fig:degen_example}
\end{figure*}

\begin{figure*}
	\centering
	{
		\includegraphics[trim= 0 0 0 0, width=0.9\textwidth]{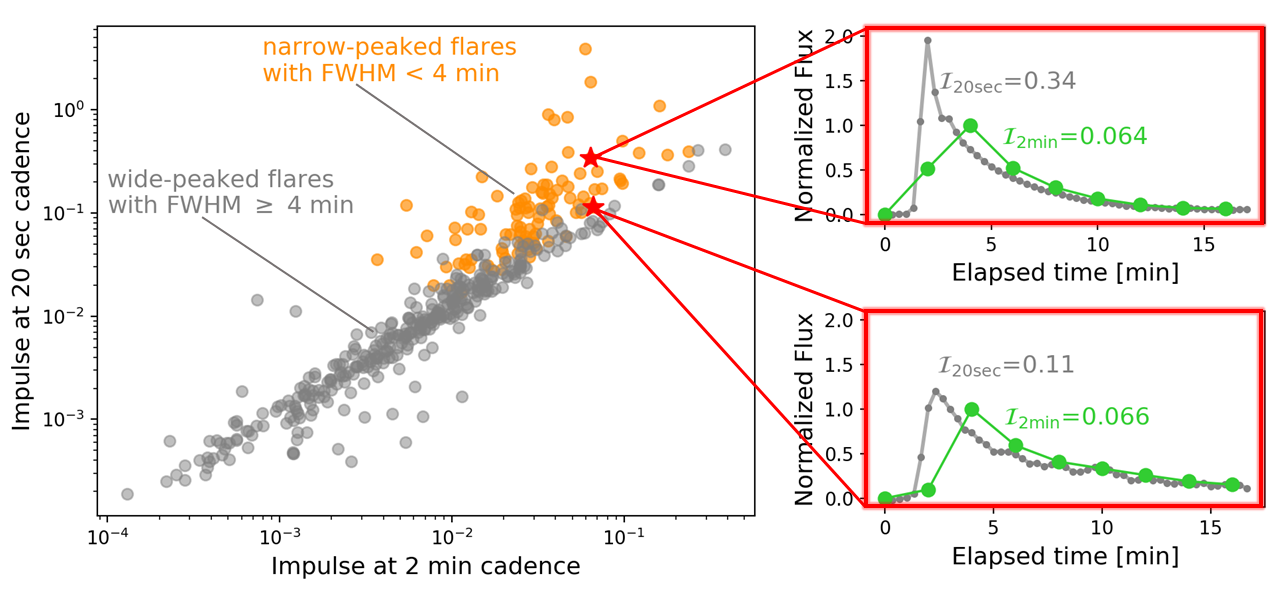}
	}
	\vspace{-0.2cm}
	\caption{Slower 2 min cadence as a predictor of flare morphology at higher 20 second cadence. Flare morphology is described by the impulse $\mathcal{I}$, which is defined as the peak flux over the FWHM duration of the flare and describes whether the peak is rapid or gradual \citep{Kowalski2013}. Left: The impulse at 2 min cadence $\mathcal{I}_\mathrm{2 min}$ does a good job of predicting the impulse at 20 second cadence $\mathcal{I}_\mathrm{20 sec}$ for the longer-lasting flares with FWHM$>$4 minutes shown in grey. The impulses of the shorter flares with FWHM$<$4min are not predicted by the two minute cadence observations as shown in orange. Right: Two call-out boxes illustrate for specific flares with the same $\mathcal{I}_\mathrm{2 min}$ how the impulse at 2 min cadence fails to predict the impulse at 20 second cadence. The green light curves are binned to 2 min cadence while the 20 second data is shown in grey.}
	\label{fig:twentysec_statistics}
\end{figure*}

Finally, we measure the parameters of each flare after the second de-trending step. Flare amplitudes are measured in the de-trended light curves using fractional fluxes instead of magnitudes. Fractional flux is computed as $\Delta$F/F=$\frac{\mid F-F_0 \mid}{F_0}$ where $F_0$ is the out-of-flare flux and is determined from the local median. We integrate the area under the fractional flux curve between the start and stop times to measure the equivalent duration (ED) in seconds. We measure the quiescent luminosity $Q_0$ of each flare star in $T$ in erg $s^{-1}$. The quiescent luminosity is computed using the method of \citet{Howard2018, Howard2019, Howard2020b} using the $T$=0 to flux calibration \citep{Sullivan2015}, the stellar distance, and the $T$ magnitude. Stellar distances and the $T$ magnitude of the star are primarily obtained from the TESS Input Catalog version 8 (TIC; \citealt{Bailer_Jones2018,Stassun2019}). Flare energies in the TESS bandpass $E_\mathrm{T}$ are computed as $E_\mathrm{T} = Q_0\times$ED. We estimate the bolometric energy $E_{bol}$ following the method of \citet{Osten2015} assuming a 9000 K flare, which produces a conversion factor of 0.19 from the bolometric to the TESS bandpass. We note that much higher optical color-temperatures have been reported (e.g. \citealt{Kowalski2010, Howard2020b}). Assuming a higher temperature would shift the dominant region of emission further into the UV and increase our estimate of the bolometric flare energy.

Flares are distributed according to flare frequency distributions where the vast majority of flares are very small and large flares are comparatively rare, e.g. \citep{hawley2014}. Since the scope of the current work is to investigate the time-resolved properties of very large flares, we cut all flares with $E_\mathrm{bol}<$10$^{32}$ erg from our sample to ease the human vetting process. Approximately 2700 candidates were discarded as $E_\mathrm{bol}<$10$^{32}$ erg events or as false positives. In the low-energy regime, a majority of the unvetted 10$^{31}$ erg candidates are likely to be false positives or indistinguishable from false positives based on the increased number of outliers in the 20 second cadence light curves. Since the 10$^{32}$ erg cut can remove high amplitude flares from M4-M6 dwarfs, we confirm only 25 large events were removed. These 25 flares exhibit comparable time-resolved properties to the remaining flares. In total, we identify 3792 real flares of $\geq$10$^{32}$ erg in the TESS light curves of the 226 stars.

\subsection{Criteria for inclusion in a well-characterized statistical sample}\label{making_stat_sample}
The signal-to-noise (S/N) of the impulsive peaks of 10$^{32}$ erg flares is much lower for early M-dwarfs than mid-to-late M-dwarfs. For example, a 10$^{33}$ erg flare may be $\sim$0.1 mag from an M1 dwarf \citep{Howard2019}, while a comparable energy flare may be $\sim$4 mag from an M5 dwarf \citep{Howard2018}. However, the scalar parameter ED (area-under-the-curve) is much more representative of the apparent size and S/N of a flare in a light curve than is the energy \citep{Loyd2018}. To mitigate the effects of uneven detection bias toward larger or smaller flares across the sample, we remove all flares too small for clear identification of sub-structure as follows:
\begin{itemize}
    \item First, we classify all 3792 flares as ``low quality" or ``high quality" during visual inspection. High quality flares are defined as sufficiently high S/N to visually identify the shape of the flare peak and whether sub-structure is present or not. Examples are shown in Figure \ref{fig:highlowquality}.
    \item Next, we compute the ED for all flares in the sample. We choose ED as the best flare parameter for identifying flares large enough to observe sub-structure since ED (a) does not change as a function of stellar mass, and (b) is a single parameter that includes contributions from both the amplitude and duration of the flares. Both a long-duration but very low amplitude flare and a very short but high-amplitude flare may be too small to clearly tell whether sub-structure is present.
    \item We compare the ED values of the events classified as low and high quality, respectively. We find that the vast majority of events flagged as ``low quality" fall below ED$\approx$50 seconds (i.e. 2479 flares in total). Therefore, this value is chosen as the cutoff for identifying large flares for further statistical analysis. Since only 12.5\% of the flares above ED$_\mathrm{cut}$ were classified as ``low quality" (i.e. 63 flares in total), removing them does not induce significant bias from human vetting.
\end{itemize}
Our final sample is composed of 440 large flares of at least 10$^{32}$ erg and with an ED$\geq$50 seconds. Of these, 428 are superflares, and 27 are superflares of $\geq$10$^{35}$ erg. Flares of $\geq$10$^{35}$ erg are amongst the largest that M-dwarfs regularly emit and have much greater effects on planetary habitability than do superflares. For example, a single 10$^{35}$ erg event is capable of pushing an entire Earth-like atmosphere out of thermochemical equilibrium \citep{Howard2019}. Ultra-high-cadence observations of $\geq$10$^{35}$ erg M-dwarf flares are so rare in the optical, however, that only $\sim$20 exist in the literature (e.g. \citealt{Jackman2019, Jackman2020, Jackman2021}). The 20 second cadence TESS sample more than doubles the available population. Both the full sample of 3792 confirmed flares and the smaller statistical sample of 440 very large flares are available as machine-readable versions of Table \ref{table:all_finds_tab} and Table \ref{table:indiv_flares_tab}.

\begin{table*}
\renewcommand{\arraystretch}{1.6}
\caption{Catalog of all 3792 Flares Observed Across 226 M-dwarfs at 20 Second Cadence During TESS Cycle 3}
\begin{tabular}{p{1.5cm} p{1.5cm} p{1.5cm} p{1.5cm} p{0.9cm} p{1.5cm} p{0.9cm} p{1.4cm} p{0.6cm} p{0.9cm} p{1.1cm} p{1.1cm}}
\hline
TIC-ID & Start & Peak & Stop & ED & log $E_\mathrm{bol}$ & $A_T$ & $\sigma_\mathrm{peak}$ & M$_*$ & $P_\mathrm{rot}$ & Noisy? \\
 & [TBJD] & [TBJD] & [TBJD] & [sec] & [erg] & [$\Delta$F/F] &  & [M$_{\odot}$] & [d] & \\
\hline
 &  &  &  &  & ... &  &  &  &  &  \\
5656273 & 2088.7327 & 2088.734 & 2088.7482 & 10 & 33.3 & 0.019 & 6 & 0.59 & 0.4312 & yes \\
5796048 & 2090.5084 & 2090.51 & 2090.5424 & 40 & 33.9 & 0.23 & 46 & 0.42 & 0.5553 & no \\
24452802 & 2180.6463 & 2180.6484 & 2180.672 & 10 & 33.0 & 0.035 & 8 & 0.18 & 1.53 & yes \\
29853348 & 2116.9378 & 2116.9385 & 2116.9575 & 30 & 33.8 & 0.049 & 7 & 0.18 & 0.7029 & no \\
29919288 & 2117.8156 & 2117.8198 & 2117.8373 & 10 & 32.8 & 0.031 & 6 & 0.18 & 4.6424 & yes \\
31740375 & 2038.3066 & 2038.3087 & 2038.3296 & 20 & 33.1 & 0.043 & 6 & 0.36 & 3.695 & yes \\
32811633 & 2195.3431 & 2195.3447 & 2195.3829 & 20 & 32.6 & 0.073 & 13 & 0.42 & no & no \\
33864387 & 2042.0784 & 2042.0852 & 2042.1245 & 70 & 34.1 & 0.088 & 17 & 0.36 & 4.62 & no \\
38586438 & 2038.8355 & 2038.8468 & 2038.8748 & 30 & 33.8 & 0.04 & 7 & 0.36 & 1.1113 & yes \\
44796808 & 2145.273 & 2145.2748 & 2145.3216 & 110 & 34.3 & 0.157 & 38 & 0.48 & 3.72 & no \\
 &  &  &  &  & ... &  &  &  &  &  \\
\hline
\end{tabular}
\label{table:all_finds_tab}
{\newline\newline \textbf{Notes.} A subset of the parameters of 3792 M-dwarf flares observed at 20 second cadence by TESS. The full table is available in machine-readable form. Columns are TIC-ID, flare start time in TESS Barycentric Julian Date (TBJD), flare peak time in TBJD, flare stop time in TBJD, equivalent duration (ED) in seconds, the log of the flare energy in erg, the peak flare amplitude in fractional flux units, the photometric S/N of the peak, the stellar mass, the stellar rotation period $P_\mathrm{rot}$ in d, and whether the photometric scatter was too noisy to reliably determine substructure properties.}
\end{table*}

\begin{table*}
\renewcommand{\arraystretch}{1.6}
\caption{Time-resolved Properties of 440 Large Flares Observed Across 226 M-dwarfs at 20 Second Cadence During TESS Cycle 3}
\begin{tabular}{p{1.3cm} p{1.3cm} p{0.7cm} p{0.9cm} p{0.9cm} p{1.5cm} p{0.7cm} p{1.3cm} p{0.8cm} p{0.9cm} p{1.2cm} p{1.7cm}}
\hline
TIC-ID & Obs. & ED & log $E_\mathrm{bol}$ & $A_T$ & $\mathcal{I}_T$ & $N_\mathrm{dom}$ & Complex rise? & QPP? & M$_*$ & $P_\mathrm{rot}$ & HZ $Flu_\mathrm{UVC}$ \\
 & [TBJD] & [sec] & [erg] & [$\Delta$F/F] & [$A_T$/FWHM] &  &  &  & [M$_{\odot}$] & [d] & [J m$^{-2}$] \\
\hline
 &  &  &  &  &  & ... &  &  &  &  & \\
5656273 & 2096.3355 & 430 & 35.2 & 0.595 & 0.08051 & 1 & yes & no & 0.59 & 0.431 & 410 \\
10863087 & 2123.8696 & 300 & 34.4 & 0.189 & 0.0098 & 3 & yes & no & 0.25 & 0.859 & 320 \\
29853348 & 2122.3386 & 120 & 34.4 & 0.185 & 0.02103 & 2 & no & no & 0.19 & 0.703 & 1310 \\
31740375 & 2050.4967 & 240 & 34.3 & 0.166 & 0.03917 & 1 & no & no & 0.36 & 3.7 & 120 \\
33864387 & 2042.0852 & 70 & 34.1 & 0.088 & 0.01045 & 1 & yes & no & 0.36 & 4.6 & 170 \\
38586438 & 2089.0641 & 110 & 34.4 & 0.118 & 0.00838 & 1 & no & no & 0.36 & 1.111 & 240 \\
44796808 & 2145.2748 & 110 & 34.3 & 0.157 & 0.02428 & 1 & no & no & 0.48 & 3.7 & 110 \\
49593799 & 2118.2676 & 330 & 33.9 & 0.595 & 0.20716 & 1 & no & no & 0.29 & no & 220 \\
49672084 & 2213.6100 & 60 & 33.6 & 0.123 & 0.05571 & 1 & no & no & 0.25 & 9.7 & 180 \\
64053930 & 2147.4245 & 70 & 34.4 & 0.088 & 0.00771 & 1 & yes & no & 0.29 & no & 530 \\
77957301 & 2193.4825 & 150 & 34.1 & 0.176 & 0.01803 & 2 & no & no & 0.20 & 1.29 & 460 \\
80427281 & 2096.1687 & 50 & 34.1 & 0.634 & 1.86471 & 1 & no & no & 0.39 & 3.8 & 1150 \\
95328477 & 2224.3931 & 110 & 34.7 & 0.204 & 0.05038 & 1 & no & no & 0.59 & 2.61 & 200 \\
117748478 & 2179.9384 & 80 & 34.2 & 0.035 & 0.00025 & 2 & no & no & 0.25 & 0.611 & 150 \\
117874959 & 2095.7303 & 1170 & 35.6 & 1.074 & 0.08295 & 1 & yes & yes & 0.49 & 6.2 & 1380 \\
 &  &  &  &  &  & ... &  &  &  &  & \\
\hline
\end{tabular}
\label{table:indiv_flares_tab}
{\newline\newline \textbf{Notes.} A subset of the parameters of 440 very large M-dwarf flaring events observed at 20 second cadence TESS. The full table is available in machine-readable form. Columns are TIC-ID, flare peak time in TESS Barycentric Julian Date (TBJD), equivalent duration (ED), the log of the flare energy, the peak flare amplitude, the impulse, the number of dominant peaks that best describe the overall shape of the light curve ($N_\mathrm{dom}$), whether rise phase complexity is present, whether QPPs are present, the stellar mass, the stellar rotation period $P_\mathrm{rot}$, and the amount of UVC fluence from the 20 second flare peak reaching the surface of a habitable zone planet in the absence of a significant atmosphere. Columns not shown but included in the machine-readable table include the flare start and stop times in TBJD, the FWHM duration in minutes, the number of substructure peaks in the rise phase ($N_\mathrm{rise}$), the number of peaks across the entire flare ($N_\mathrm{tot}$), the log of the quiescent luminosity $Q_0$, and the estimated survival fractions of common micro-organisms after 20 seconds of peak UVC emission.}
\end{table*}

\section{Exploring flare morphology at 20 second cadence}\label{2min_vs_20sec}

To ensure accuracy, we visually classify each of the 440 flares in our final sample into classical and complex flares, recording the number of peaks from each flare. We record the number of peaks that best describes the general shape of the flare light curve $N_\mathrm{dom}$, the total number of peaks $N_\mathrm{tot}$, and the number of peaks clearly visible in the rise phase $N_\mathrm{rise}$. The number of peaks in the decay phase may be obtained by subtracting $N_\mathrm{rise}$ from $N_\mathrm{tot}$ values in Table \ref{table:indiv_flares_tab}. These three numbers generally correlate well with each other. We find $N_\mathrm{dom}$ to be the most useful for flare classification and least susceptible to noise. At 20 sec cadence, there is enough substructure it is helpful to distinguish between the overall shape of the flare light curve and the small perturbations to that overall shape. This is what $N_\mathrm{dom}$ and $N_\mathrm{tot}$ attempt to do. $N_\mathrm{dom}$ is a visual classification metric of whether the large-scale structure of the flare is ``best described by" one peak or multiple peaks. $N_\mathrm{tot}$ is the total number of impulsive spikes of any size in the flare light curve. These values are recorded in Table \ref{table:indiv_flares_tab}. Within the sample, 57.7\% of flares show a classical morphology with a single dominant peak, 31.8\% have two dominant peaks, 8.2\% have three dominant peaks, and 2.3\% have four or more such peaks. In total, we observe 42.3\% complex flares. The complex flare rate is higher at the higher energies; 70$\pm$9\% of flares above 10$^{35}$ erg are complex. The breakdown in number of peaks we report is broadly consistent with the relative numbers of complex peaks from 3737 GJ 1243 flares obtained by \citet{Davenport2016proc} with 1 minute cadence \textit{Kepler} observations.

\subsection{Can 2 min cadence predict higher cadence behavior?}\label{comparing_cadences}
The new 20 second cadence mode reveals significant substructure in large flares that would have been missed at 2 min cadence. Higher-cadence observations also remove degeneracy present at 2 min cadence between significantly different flare morphologies. A particularly compelling example of the ability of the 20 second cadence observations to distinguish between different morphologies is shown in Figure \ref{fig:degen_example}. While both flares appear similar at 2 minute cadence, the flare on the left is resolved at 20 second cadence as having an extended period of near-constant emission at peak while the flare on the right has a classical impulsive profile.

\begin{figure}
	\centering
	\subfigure
	{
		\includegraphics[trim= 0 50 180 0, width=0.5\textwidth]{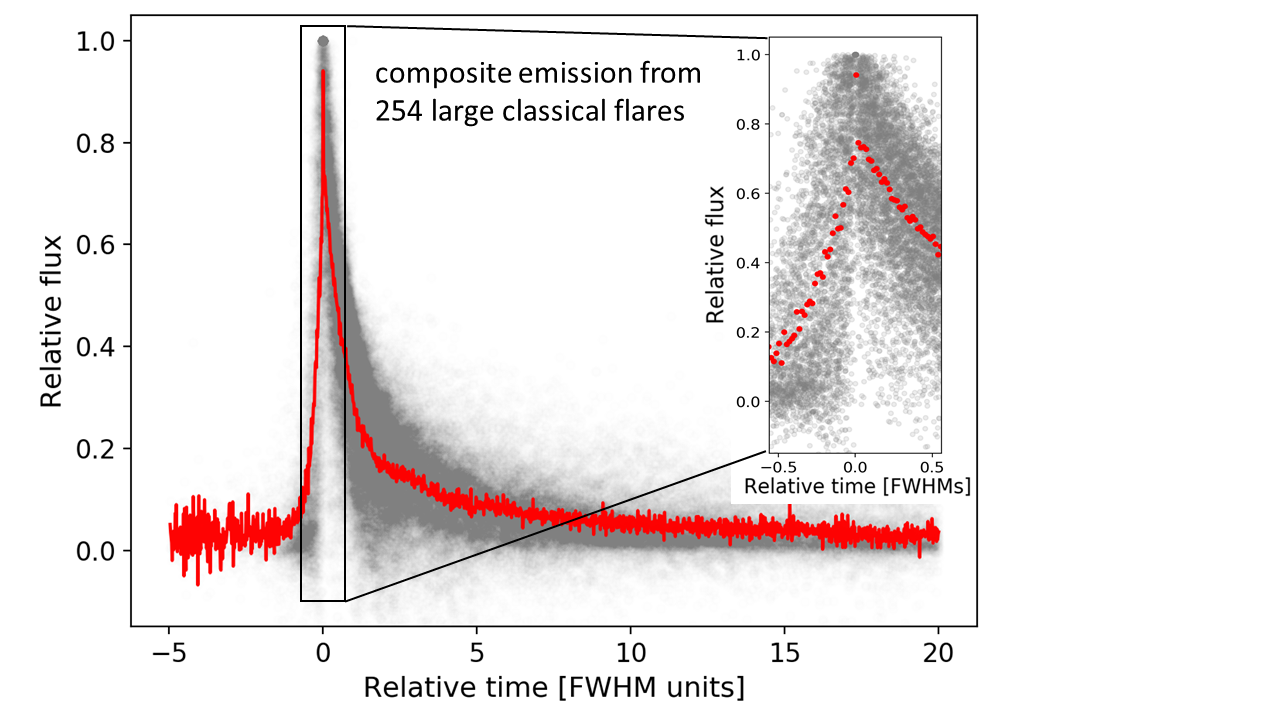}
	}
    \subfigure
	{
		\includegraphics[trim= -10 0 -15 0, width=0.5\textwidth]{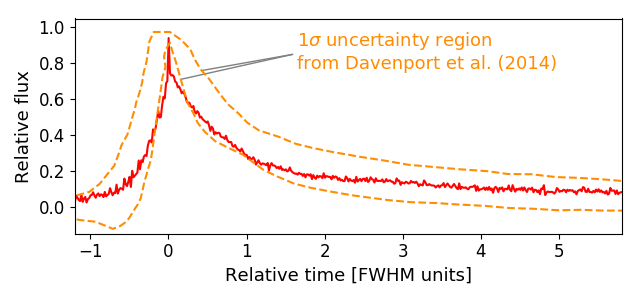}
	}
	\vspace{-0.2cm}
	\caption{Top: A ``composite flare" composed of 254 classical flares obtained at 20 second cadence. Flares are normalized in both amplitude and time (by dividing by the FWHM time of each flare). Our composite flare strongly supports the overall validity of the original \citet{davenport2014} flare template shown in their Figure 4, but with two orders of magnitude more M-dwarfs and 3$\times$ higher cadence. A cutout of the resolved composite peaks is shown as an inset. It is unclear if the rounded peak in the inset is astrophysical or due to imperfect overlay of the constituent flares. We also note non-typical flare peak morphologies are suppressed by the averaging process. Bottom: The 1$\sigma$ uncertainty region from Figure 4 of \citet{davenport2014} is compared against our template. Other than within 0.2 FWHMs of the peak, our model lies within the uncertainties from their work.}
	\label{fig:test_of_davtemplate}
\end{figure}

The ability of the 2 minute data to predict higher-cadence behavior can be determined more broadly by comparing the morphologies of all 440 flares at both cadences. The impulse $\mathcal{I}$ serves as a good proxy for the flare morphology by measuring how rapidly the flare energy is released \citep{Kowalski2013}. We compare the impulse of each flare at 2 minute cadence and again at 20 second cadence. The results of the $\mathcal{I}_\mathrm{2 min}$-$\mathcal{I}_\mathrm{20 sec}$ comparison are plotted in Figure \ref{fig:twentysec_statistics}. The flares are color-coded based on the duration of the rapid phases of the flares given by their FWHM in minutes. While the 2 minute cadence data does predict the impulsiveness of the flares at 20 second cadence for flares with FWHM$>$4 min, this is not true for the shorter events. Figure \ref{fig:twentysec_statistics} demonstrates that any two flares of FWHM$<$4 min that have comparable impulses at 2 minute cadence can differ by more than an order of magnitude at 20 second cadence. Clearly, high cadence observations are needed to properly assess the morphologies of these events.

\subsection{The ``typical" flare at 20 second cadence}\label{testing_davenport}
While the peaks of flares observed at non-optical wavelengths often have light curves shaped like a Gaussian (e.g. \citealt{MacGregor2021}), this is not thought to be the case in the optical. Optical flares are described by a rapid-rise, exponential decay template empirically determined by \citet{davenport2014}. This template or ones derived from it are routinely used to fit flare light curves (e.g. \citealt{Vida2017, Schmitt2019, Gunter2020, Glazier2020}). This is especially important for flares where only a few epochs are observed at high signal to noise \citep{Howard2020b} or with very low cadence \citep{Schmitt2019} as the flare's energy can be accurately estimated even from sparse information. Because the \citet{davenport2014} template was constructed from 885 classical flares observed by \textit{Kepler} at 1 minute cadence from GJ 1243, our new 20 second cadence observations of 243 classical flares emitted from two orders of magnitude more M-dwarfs provide a useful test of the model.

We recreate Figure 4 of \citet{davenport2014} with our 20 second cadence observations. Following their procedure, we normalize the flux of each classical flare to have an amplitude of one. We also normalize the duration of each classical flare by its FWHM time, placing all flares onto the same time scale. We ignore the clearly complex flares for the moment in order to investigate the behavior of the 243 classical flares with single peaks. This is because we desire to test the behavior of impulsive events and it is difficult to disentangle the individual impulsive events from complex flares. We treat several cases of complex emission in Section \ref{common_morphologies}. Next, we bin the flare to observe the ``typical" flare seen at 20 second cadence. The result is shown in Figure \ref{fig:test_of_davtemplate} and is nearly identical to the original Figure 4 of \citet{davenport2014}. The composite flare emission is plotted in grey, and the binned flare is shown in red. Since the 1 minute \textit{Kepler} flares were not fully resolved in the rise phase, we also plot an inset showing the emission during this phase. We note the high scatter during the rapid phase and peak may be due to either rounded peak morphologies or imperfect alignment of the time axes using the FWHM.

Given the applicability of the \citet{davenport2014} template, we fit the peaks of each of the 243 classical flares with both the template and a generic Gaussian. For each flare, we isolate the flare peak by selecting the flare times and fluxes above the half-maximum and fit the two models. Each flare is selected to have at least 4 measurements within the FWHM to prevent indeterminate fits. 

We record the residuals from both model fits for each flare and compare the results. Although the \citet{davenport2014} model was developed from 1 min cadence data, it does an excellent job fitting the morphology of classical flares even at 20 second cadence. In fact, only 8.6$\substack{+2 \\ -1.6}$\% of classical flares are better fit by a Gaussian. Visual inspection of these few events show some to be truly Gaussian in shape while others are underfit or have substructure that is not fit well by a single exponential peak. The 8 flares with the clearest Gaussian peaks were among the highest amplitude and most impulsive flares across our larger sample, suggesting energetic electron acceleration plays a key role in forming non-Gaussian peaks. Since non-optical flares with Gaussian peaks clearly trace the initial reconnection event \citep{MacGregor2021} while optical flares are prompt emission from heating events \citep{Kowalski2013} as indicated by the exponential decay tail, perhaps the prompt emission in Gaussian-shaped optical flares reflects continuing electron acceleration and delayed cooling. One caveat is that since fits are done within the FWHM and not only at the true peak, even flares better fit by a \citet{davenport2014} template than a Gaussian template may still have non-impulsive tops such as the flare to the lower right of Figure \ref{fig:twentysec_statistics}. We also note the Gaussian-top flares are not included in the flares with constant/level peaks from Section \ref{flat_tops} since the Gaussian peak emission is not constant. 

\begin{figure*}
	\centering
	{
		\includegraphics[trim= 0 0 0 0, width=0.9\textwidth]{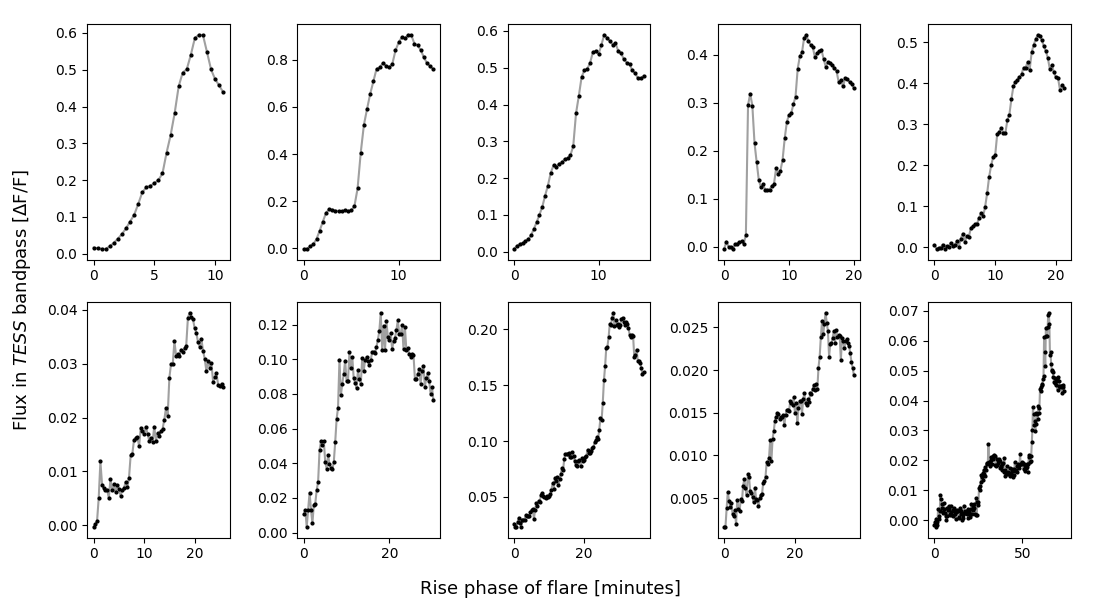}
	}
	\caption{Nearly half of the flares in our sample show complex substructure during the rise phase. We display a subset of 10 flares showing rise phase complexity here. A greater degree of complexity generally correlates with longer rise times, although exceptions exist. Substructure in the rise phases of large M-dwarf flares is difficult to resolve in lower-cadence observations.}
	\label{fig:rise_phase_complexity}
\end{figure*}

\subsection{Complex emission in the rise phase}\label{rise_phase}
Many large flares exhibit complex substructure during the rise phase. We define the rise phase to be the time between the beginning of a flare and when it reaches its peak brightness. Because slower cadences often fail to resolve the rise phase \citep{Kowalski2019}, our sample provides a testbed to explore the properties of the rise phases of large flares. Complex emission in the rise phase has been previously observed from both the Sun and other stars (e.g. \citealt{Kosovichev2001,Veronig2010,Kowalski2019, Tamburri2021}). On the Sun, rise phase complexity is often due to spatio-temporal flare evolution. Accelerated electrons are thought to heat the lower atmospheric layers in adjacent emission regions, but not simultaneously or with the same intensity \citep{Veronig2010, Tamburri2021}. During the 2000 Bastille Day solar flare \citep{Kosovichev2001}, emission from two separate but interconnected two-ribbon flares evolved along a polarity inversion line to produce an optical light curve with two peaks (e.g. \citealt{Qiu2010, Kowalski2019}). This mechanism has also been proposed by \citet{Kowalski2019} to explain complex emission in the rise phase observed from an M-dwarf stellar flare from GJ 1243.

We find 46\% of the large flares in our sample exhibit complex structure in the rise phase (201 out of 440 flares), making this a common phenomenon at 20 second cadence. We note that flares classified earlier as having broadly classical morphologies can also have rise phase complexity since these estimators are for different properties of the flare structure. Rise phase complexity allows for minor peaks and inflection points in the rise phase, while the number of ``classical" flares is the number of flares whose large-scale light curve morphology is best described by one \citet{davenport2014} template rather than a superposition of templates.

Among flares with rise phase complexity, more substructure in the rise phase correlates with longer rise phases. While not a one-to-one relationship, flares with higher $N_\mathrm{rise}$ values are statistically characterized by longer rise times. An Anderson-Darling (A-D) test of the distribution of rise times containing one substructure component and the distribution of longer rise times with two components is able to distinguish between them to a p-value of $p<$0.01. Likewise, an A-D test of the rise times containing two substructure components and the distribution of longer rise times with three or more components is able to distinguish between them to a p-value of $p<$0.01. There are too few flares to test higher orders separately. 13\% of our full sample (57 out of 440 flares) shows a double-peaked structure reminiscent of the Bastille Day flare and the GJ 1243 M-dwarf flare \citep{Kowalski2019}. Examples of rise-phase complexity from a subset of 10 large flares are shown in Figure \ref{fig:rise_phase_complexity}.

\begin{figure*}
	\centering
	{
		\includegraphics[trim= 0 20 0 0, width=0.9\textwidth]{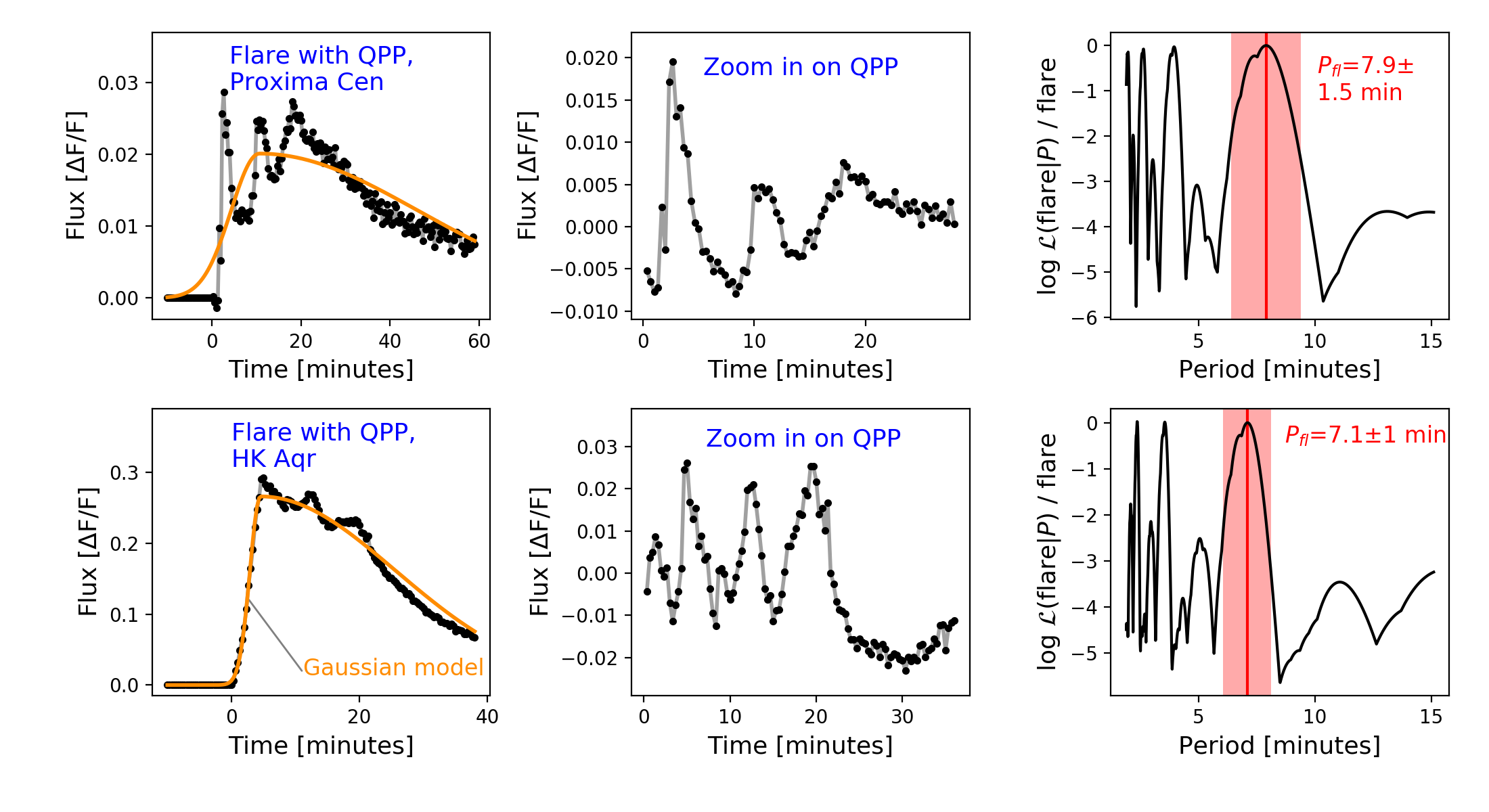}
	}
	\caption{Characterization process for quasi-periodic pulsations (QPPs) in 20 second TESS data, illustrated for a rise-phase QPP from Proxima Cen and a decay-phase QPP from HK Aqr. We fit and remove a two-component Gaussian flare model in the left panels to isolate the substructure. The two Gaussians must have the same central positions and heights but the widths are allowed to differ. This is done to account for the rapidity of the rise phase of the flares. The residual substructure after the model is subtracted is shown in the middle panels. The right panels show the period and period error measurement process. A Bayesian log-likelihood flare periodogram is computed and the highest peak consistent with the substructure component inter-arrival times is selected. Period errors are measured as the FWHM of the selected peak. Each peak is also confirmed with LS periodograms of the light curve variability from the middle panel.}
	\label{fig:qpp_methods}
\end{figure*}

\begin{figure*}
	\centering
	{
		\includegraphics[width=0.98\textwidth]{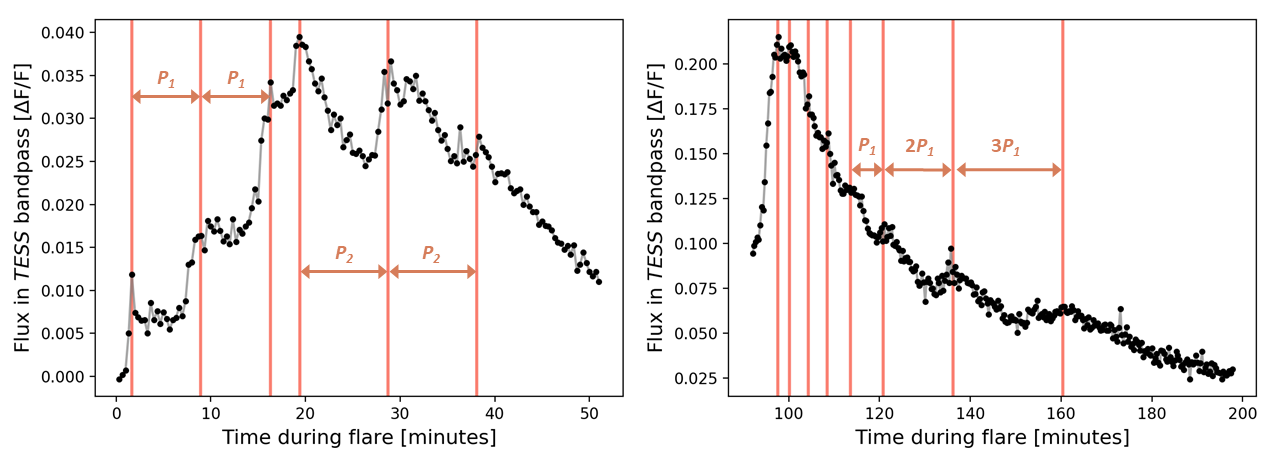}
	}
	\caption{Two examples of candidate QPPs showing evidence of period evolution. Substructure in the left flare is consistent with a period of 7.32 min in the rise phase, but a slightly longer period of 9.34 min during the decay phase. The ``period" in the right flare constantly increases from 4.13 min up to $\sim$22 min. However, the last three events show a harmonic spacing of 7.3 min, 2$\times$ 7.3 min, and 3$\times$ 7.3 min indicative of wave-like modes.}
	\label{fig:multimodal}
\end{figure*}

\begin{figure*}
	\centering
	{
		\includegraphics[trim= 0 0 0 0, width=0.84\textwidth]{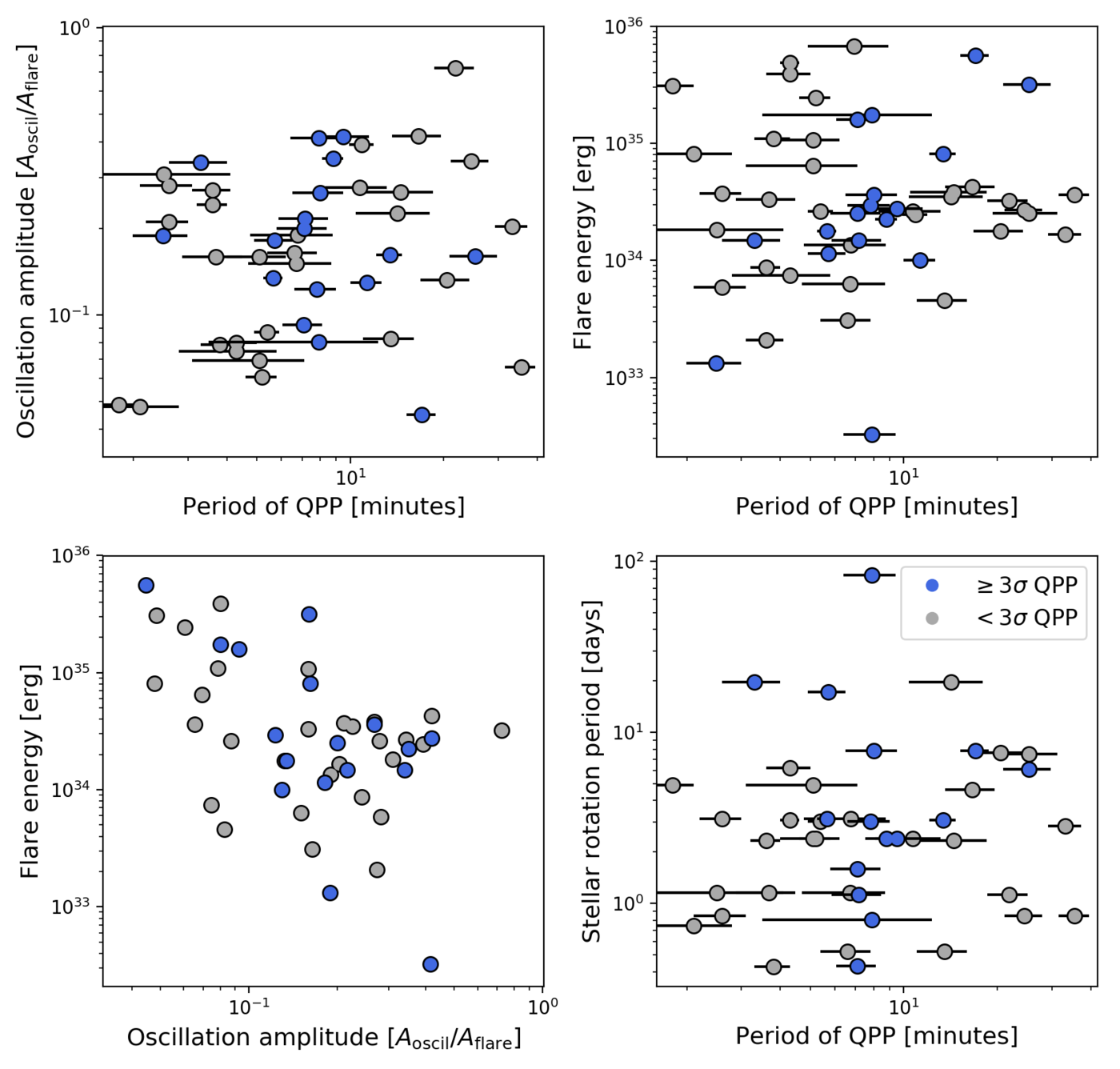}
	}
	\caption{Relationships between QPP period, oscillation amplitude, and stellar flare parameters. Candidate QPPs are shown in grey while confirmed 3$\sigma$ QPPs are shown in blue. Across all four panels, a clear trend is present in only the oscillation amplitude versus flare energy plot (lower left). A full description of these relationships is given in Section \ref{qpp_flares} of the main text.}
	\label{fig:qpp_relations}
\end{figure*}

\section{A 20 Second Cadence View of Quasi-periodic pulsations}\label{qpp_flares}

It is difficult to measure short-period QPPs of $\sim$10 min or fewer in stellar flares with cadences of 1 min or slower (e.g. \citealt{Balona2015, Ramsay2021, Monsue2021}). This observational bias has hampered statistical studies of short-period QPPs in optical stellar flares. Our statistical sample of 440 flares observed at 20 second cadence enables us to compare the properties of short-period QPPs with longer-period ones.

We tag 49 flares as candidate QPPs in our sample during an initial visual search for apparent periodicity in the flare substructure. Possible periodicity was noted whenever substructure appeared to undulate or showed regular increases in emission. As with automated discovery, the decaying variability amplitudes limited sensitivity to long period oscillations relative to short periods. We characterize each candidate as follows:
\begin{itemize}
    \item The flare light curve is fit with a two-component Gaussian that is joined at the peak time of the flare. On either side of the peak, a different width is allowed by the fit to account for faster rise times than decay times. This process is shown in the first column panels of Figure \ref{fig:qpp_methods}. The model fit is subtracted from the flare light curve to isolate the QPP substructure as shown in the middle column panels of Figure \ref{fig:qpp_methods}.
    \item We identify the most likely period and period errors with a Bayesian flare periodogram developed in \citet{Howard_Law_2021} to determine periodicity in a discrete set of events in noisy and incomplete data. The log-likelihood of a given periodicity $P$ on a set of event times $F_i$, log $\mathcal{L}$($P | F_i$), is determined using Baye's theorem and a probability comb Prob($F_i | P_\mathrm{comb}$) that allows for slight imperfections in the arrival times of each event as described in \citet{Howard_Law_2021}. The closest spacing of inter-arrival times between each peak in the substructure and the total length of the flare set the respective lower and upper limits on the region of the periodogram to be searched. Within this period window, the largest peak is selected to determine the correct QPP period. Period errors are given by the FWHM of the selected periodogram peak. An example of this process is shown in the third column panels of Figure \ref{fig:qpp_methods}.
    \item Next, the Lomb-Scargle (LS) periodogram is computed on the subtracted light curves of the middle column panels of Figure \ref{fig:qpp_methods}. The LS power enables each candidate to be statistically confirmed or rejected in Monte-Carlo (MC) trials. The Bayesian method is a more sensitive QPP detection tool because all peaks in decaying-amplitude oscillations are weighted equally, while LS periodograms are more sensitive to higher-amplitude oscillations than smaller ones. However, it is easier to reproduce peaks with randomized event times in Bayesian periodograms than in LS periodograms since only the event times are considered.
    \item Randomized LS power is measured across 1000 MC trials. In each MC trial, the light curve is randomly shuffled and wide Gaussian bumps of $\sim$5 minutes duration are injected onto the shuffled light curves at random times to mimic substructure. An equivalent number of bumps to the actual number of substructure peaks are injected in each trial. The false positive percentage $FP$ is measured as 100 times the number of trials with random power equal or exceeding the actual signal power over the total number of trials.
\end{itemize}
Of the 49 candidates, 17 are confirmed as $\geq$3$\sigma$ detections at $FP\leq$0.3\%. The rest remain unconfirmed candidates of varying degrees of likelihood. Several of the unconfirmed QPPs show evidence of multi-modal periodicity or period evolution over the course of the flare, splitting their LS power and reducing their signal strength. Such multi-modal periodicity and period evolution is a well-known phenomenon (e.g. \citealt{Doorsselaere2016} and references therein). Two examples of our multi-period events are shown in Figure \ref{fig:multimodal}. Some QPPs are observed in the rise phase instead of the decay phase, such as the one from Proxima Cen in Figure \ref{fig:qpp_methods}.

Candidate QPPs in our sample have periods ranging from 2 to 36 minutes, and confirmed QPPs have periods ranging from 3 to 25 minutes. Previously published M-dwarf QPPs from 2 minute cadence TESS data have periods as short as 10.2 minutes \citep{Vida2019, Ramsay2021}. \citet{Million2021} report 9 QPPs seen in unpublished 20 sec cadence TESS data, with periods ranging from 2 to 8.7 minutes. Most QPPs observed in 1 minute cadence \textit{Kepler} data generally have periods of $\sim$10 minutes or longer \citep{Pugh2016}. 33 of our 49 candidate QPPs and 13 of our 17 confirmed QPPs have periods less than 10 minutes. The relative lack of QPPs at short periods seen in previous statistical studies are therefore likely a result of the observing cadence. QPP periods of 1 minute cadence \textit{Kepler} flares from \citet{Pugh2016} are consistent with a uniform distribution of periods from 10 to 50 minutes. Such a consistent detection efficiency may be a result of the lower photometric scatter at 1 min cadence. If so, 2 minute cadence TESS data may produce an increased number of long-period QPPs. We note our detection efficiency for long-period QPPs may be reduced relative to short period ones as a result of higher photometric scatter and decaying amplitudes.

In Figure \ref{fig:qpp_relations}, we explore relationships between the period and relative oscillation amplitude of our QPPs with each other and with other parameters. We define the relative oscillation amplitude $A_\mathrm{oscil}/A_\mathrm{flare}$ as the ratio of the peak-to-trough flux amplitude of the second-largest feature in the flare over the peak flux of the flare itself. We find this normalization relative to the size of the flare produces significantly less scatter in the relationships shown here than does the raw flux of the QPP features themselves. In the top left panel, we show the QPP period versus the oscillation amplitude. A possible trend of larger amplitudes at longer periods may exist, but it is tentative at best. The top right panel shows the QPP period versus the flare energy. Longer periods show an increasing floor in the minimum energy and a decreasing scatter in energy compared to shorter-period QPPs that may be due to reduced detection efficiency at longer periods. This results from long period QPPs only being detectable in higher energy, longer duration flares. Since the duration of the flare tail depends on the photometric scatter of the light curve, it is likely the increased noise of the 20 second cadence light curves leads to reduced efficiency in detecting long period QPPs. The bottom left panel shows higher-energy flares have smaller relative oscillation amplitudes. The bottom right panel shows no correlation between QPP period and stellar rotation period. Rotation periods were measured from high-amplitude rotational variability in the quiescent SAP fluxes using LS periodograms and confirmed by eye. The trend of longer QPP periods at longer stellar rotation periods in \citet{Pugh2016} exists at longer QPP periods than are present in our sample.

\section{Other unusual but frequently-occurring flare morphologies}\label{common_morphologies}

\subsection{Peak-bump flares}\label{peak_bump_flares}
A particularly notable and recurring flare morphology is described by an initial and highly impulsive event followed by a much less impulsive (or even Gaussian) secondary event. Approximately 17\% of all complex flares in our sample exhibit this ``peak-bump" profile. While some of these peak-bump flares are explainable as a random superposition of two sympathetic flares with typical fast-rise, exponential-decay (FRED) profiles, most peak-bump events are probably causally related by a two-phase underlying emission mechanism. Such flares have been recorded before (e.g. \citealt{Gunter2020, Jackman2021}) but have not yet been systematically explored in large samples at high cadence.

\begin{figure*}
	\centering
	{
		\includegraphics[trim= 0 0 0 0, width=\textwidth]{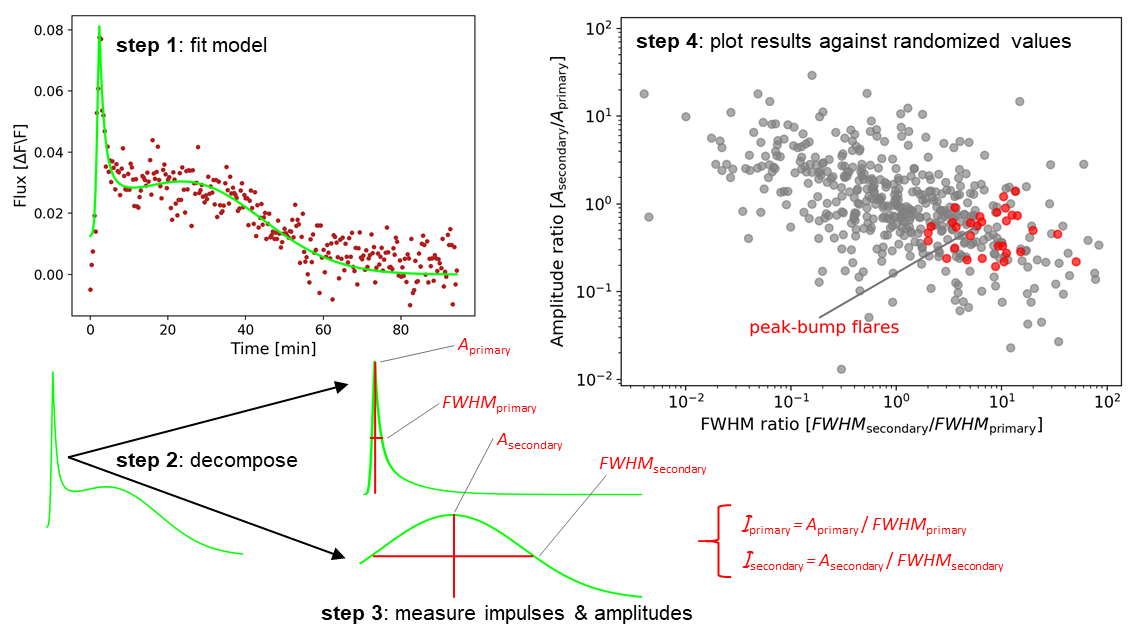}
	}
	\vspace{-0.5cm}
	\caption{Complex flares exhibiting the peak-bump morphology appear to cluster together in impulse-amplitude space. To measure the ratio of FWHM widths, amplitudes, and impulses between the first and second peaks of the flares, we first fit them with a model composed of a \citet{davenport2014} flare template and a Gaussian (top left). This allows the primary and secondary components to be separated and accurate amplitudes and FWHM to be measured (bottom left). The resulting ratios of the secondary to the primary components demonstrate flares that were classified as ``peak-bump" routinely display a period of highly impulsive emission followed by a much less impulsive follow-on event. For comparison, randomized FWHMs and amplitudes from our larger sample of flares are divided by one another and plotted in grey against the peak-bump flares in red (right).}
	\label{fig:peak_bump_methods}
\end{figure*}

\begin{figure*}
	\centering
	{
		\includegraphics[trim= 0 0 0 0, width=0.99\textwidth]{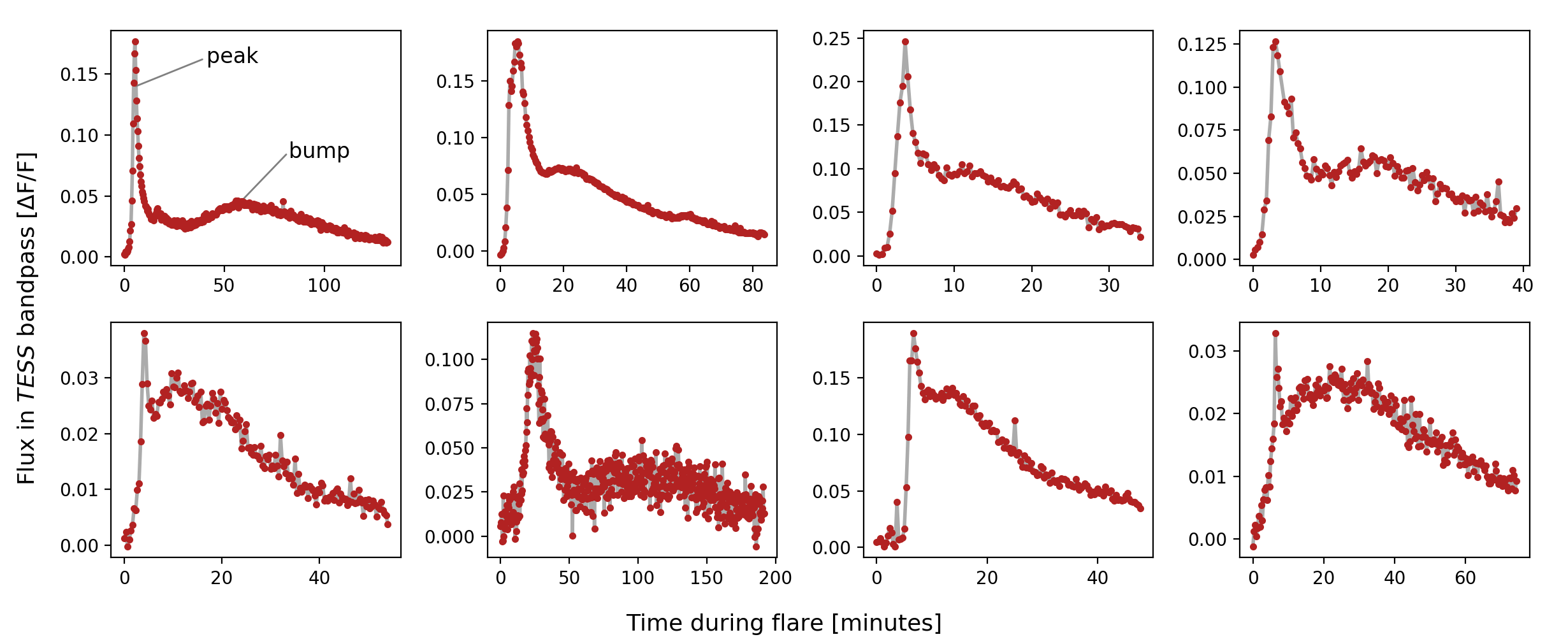}
	}
	\caption{Examples of ``peak-bump" flare morphologies. Peak-bump flares are characterized by a large, highly-impulsive flare peak followed by a lower-amplitude and less impulsive Gaussian peak. 26\% of our complex flares show a rapid spike prior to a period of complex flare emission. We hypothesize these events may reflect an initial reconnection event that overcomes a large potential barrier to rapidly release stored energy, followed by a more gradual release of energy.}
	\label{fig:peak_bump_examples}
\end{figure*}

The emission of peak-bump flares is well-described by a model composed of a \citet{davenport2014} template and a Gaussian as shown in Figure \ref{fig:peak_bump_methods}. The secondary Gaussian peaks are not included in the statistics of classical flares with Gaussian peaked-events since these flares are complex and it is difficult to tell how secondary events in complex flares are related to single peaks in classical flares. Fitting this model to our peak-bump events allows us to decompose the two phases of the flares into separate events. We therefore separately measure the amplitude and FWHM of each component peak to obtain the impulsiveness of both components. We then place peak-bump flares into a quantitative framework with three quantities: (1) the ratio between the amplitude of the primary and secondary peaks $\mathcal{R}_A = A_\mathrm{secondary}/A_\mathrm{primary}$, (2) the ratio between the widths of the primary and secondary peaks $\mathcal{R}_\mathrm{FWHM} = FWHM_\mathrm{secondary}/FWHM_\mathrm{primary}$, and (3) the ratio between the impulsiveness of the primary and secondary peaks $\mathcal{R}_\mathcal{I} = \mathcal{I}_\mathrm{secondary}/\mathcal{I}_\mathrm{primary}$. We place the second component on top so that smaller ratios correspond to smaller and less impulsive secondary events relative to the primary.

The ratios $\mathcal{R}_A$, $\mathcal{R}_\mathrm{FWHM}$ and $\mathcal{R}_\mathcal{I}$ together determine the properties of peak-bump flares and help quantify the chance that peak-bump flares are due to random superposition and not a unifying underlying physical mechanism. Flares exhibiting the peak-bump morphology have a mean and 1$\sigma$ range of $\mathcal{R}_A$ = 0.5$\pm$0.3, $\mathcal{R}_\mathrm{FWHM}$ = 10.1$\pm$9.7, and $\mathcal{R}_\mathcal{I}$ = 0.09$\pm$0.07, respectively. The likelihood that peak-bump flares are due to the random occurrence of two sympathetic flares can be tested against randomized $\mathcal{R}_A$ and $\mathcal{R}_\mathcal{I}$ values obtained from the distributions of FWHMs and amplitudes measured for all 440 flares in Table \ref{table:indiv_flares_tab}. Because impulse is dependent on the other two quantities, it is less useful for 2D comparisons. We plot the randomized ratios in grey in the top right panel of Figure \ref{fig:peak_bump_methods} and the actual distribution of peak-bump flares in red. Next, the number of randomized flares with FWHM ratios greater than the mean $\mathcal{R}_\mathrm{FWHM}$ value and amplitude ratios less than the mean $\mathcal{R}_A$ value is divided by the total number of randomized flares. We find that randomized flares can reproduce a peak-bump morphology 5-8\% of the time. We would therefore expect 15$\substack{+3 \\ -6}$ peak-bump flares from chance in a sample of 186 complex flares. However, we observe $\sim$31 peak-bump flares in the data (and more with pseudo-peak-bump shapes), a $\geq$3$\sigma$ discrepancy.

Furthermore, all peak bump flares cluster tightly in $\mathcal{R}_\mathcal{I}$. In fact, 10,000 Monte Carlo (MC) draws of 31 flares from the randomized distribution cluster as tightly as the real 31 peak-bump flares toward small values only 0.2\% of the time. To simulate the effects of human bias in the selection of the peak bump sample prior to measuring their $\mathcal{R}_\mathcal{I}$ values, all MC draws are constrained to have $\mathcal{R}_\mathcal{I}<1$ so that the first component is more impulsive than the second. Because the true selection function of human classification cannot be known, we caution the reader in over-interpreting these results.

We hypothesize peak-bump morphologies result from an initial reconnection event that overcomes a large potential barrier to rapidly release stored energy. Once the potential barrier has been overcome, a more gradual subsequent reconnection event (or cascade of smaller events) can occur. The more gradual ``bump" phase might plausibly reflect changes in flare heating with time, such as emission from prolonged, low-level heating during the decay phase of the primary event \citep{Kowalski2013} or decreasing color-temperatures as the dominant source of heating transitions from the base of the stellar atmosphere up into the corona \citep{Kowalski2013}. If true, then reduced electron heating levels during the bump may lead to lower FUV and millimeter fluxes than during the primary peak as these wavelengths trace electron acceleration \citep{MacGregor2021}. Supporting this hypothesis, \citet{Kowalski2016} obtained APO/DIS low resolution spectra of several flares with peak-bump morphologies and found the temperatures of the gradual bump phase were consistently $\sim$3000 K lower than the initial event. We also note $\sim$9\% of complex flares in our sample not classified as ``peak-bump" also exhibit a highly-impulsive event prior to the release of the majority of flare energy. Together, $\sim$26\% of all complex flares show either a peak-bump shape or a highly-impulsive spike at the beginning of a period of complex flare emission.

\subsection{Flat-top flares}\label{flat_tops}
While flares are typically modelled with a strongly impulsive peak (e.g. the \citet{davenport2014} flare template), we observe occasional ``flat-top" flares with constant or relatively constant emission levels at peak (see the example in Figure \ref{fig:degen_example}). ``Flat-top" flares still exhibit a rapid rise and exponential decay outside of the peak phase. Such flares have been previously reported with peaks lasting for minutes up to nearly an hour. Solar flares exhibiting a flat top morphology have been observed in both the X-ray \citep{BakSteslicka2010} and in white light emission \citep{Hudson2006}. In the low mass context, \citet{Jackman2021} published a NGTS M-dwarf flare with the ``flat-top" morphology that lasted only minutes and would not have been detected in 1 min \textit{Kepler} or 2 min TESS light curves. \citet{Jackman2019} reported another NGTS M-dwarf flare with QPPs during the peak that levelled off for nearly an hour before decaying. QPPs that occur during the flare peak are therefore one mechanism that can prolong the peak phase. Finally, \citet{Romy2018} reported a $\sim$10 magnitude mid M-dwarf flare in V band by ASAS-SN that showed constant emission at peak for at least $\sim$5 min.

\begin{figure*}
    \vspace{-0.35cm}
	\centering
	{
		\includegraphics[trim= 0 0 0 0, width=0.99\textwidth]{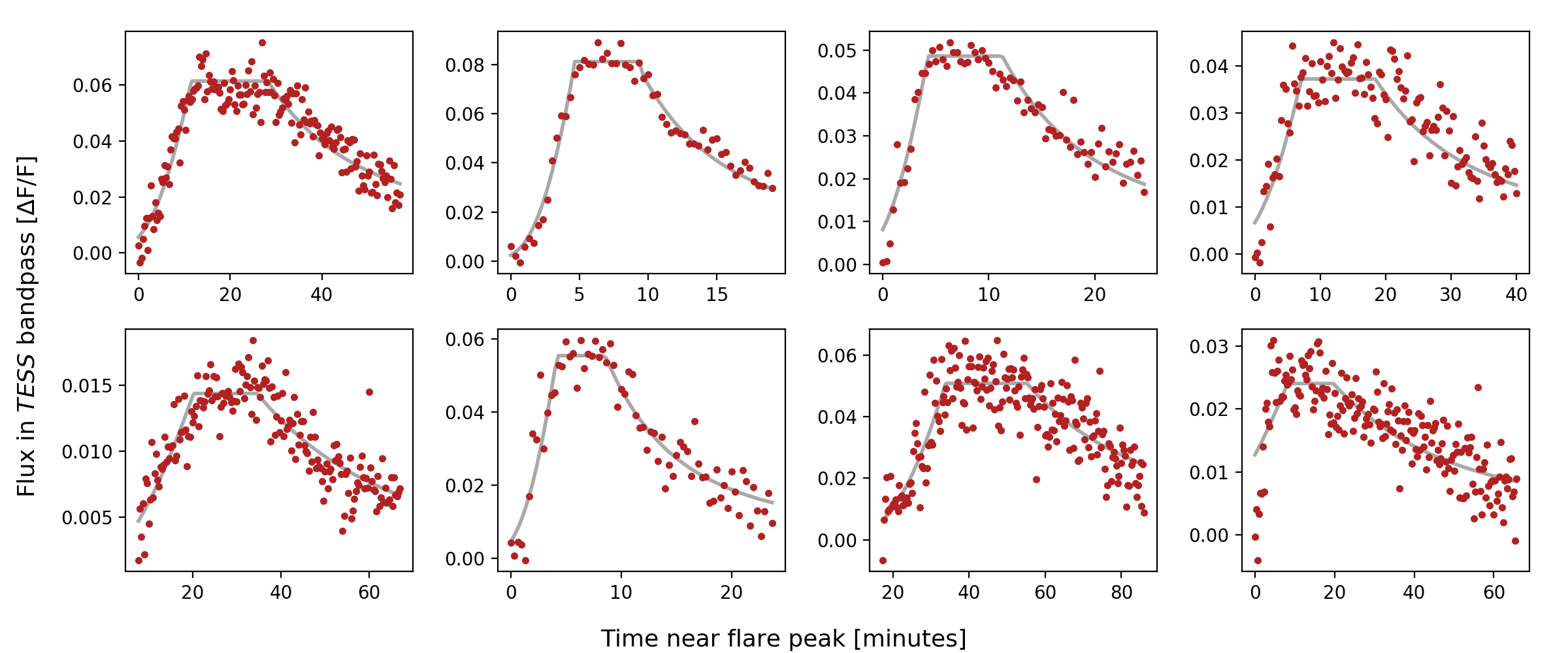}
	}
	\caption{Examples of ``flat-top" flare morphologies. While flares are often assumed to have highly impulsive peaks, high-cadence observations show some flares emit at relatively constant levels once reaching peak brightness before decaying again. We fit \citet{davenport2014} flare templates with the peak flux replaced by a constant value, shown in grey. The bottom right flare may be composed of multiple unresolved events at peak.}
	\label{fig:flat_top_flares}
\end{figure*}

We classified 24 flares best described as ``flat-top" events, 5\% of our total sample. Eight particularly clear examples of the flat-top morphology are shown in Figure \ref{fig:flat_top_flares}. The nearly-constant peak levels last from 2 to 26 minutes before beginning the decay phase. Several of these flares are clearly a superposition of unresolved multi-peaked events. Others may be unresolved QPPs (e.g. \citealt{Jackman2019}). The rest likely result from a prolonged emission event. Since flat-top flares often exhibit a prolonged period of emission at peak, they are among the less impulsive flares in our sample. Flat-top events generally occur uniformly across the left side of the grey distribution of Figure \ref{fig:twentysec_statistics}. Because flat-top flares do not have typical emission profiles, it is likely that the temporal evolution of particle and UV emission from these flares might differ from that of classical flares. Specifically, an extended period of high electron acceleration and UV emission may occur at peak.

\section{Habitability and Astrobiological Implications of Time-resolved flare Emission}\label{hab_implication}

Habitability impacts of optical stellar flares are usually estimated from the integrated energy of the flare (e.g. \citealt{Howard2018, Gunter2020, Glazier2020}). However, the time-resolved properties of flares can provide more detailed constraints on planetary atmospheres \citep{Segura2010, Loyd2018} and surface habitability \citep{Ranjan2017, OMalley2019, OMalley2019_bio}. The peak emission of flares does not last long compared to the overall duration of flare emission, but the high fluxes reached at peak may have an outsized impact on surface habitability. Very few M-dwarf superflares have been observed at high cadence directly in the UV \citep{Howard2020b}. For example, 1904 flares were observed at 10 sec cadence in GALEX data \citep{Brasseur2019}. However, only $\sim$10 of these flares were emitted by sources identified as M-dwarfs. Intriguingly, our results in the TESS bandpass show a broad range of flare morphologies beyond the classical similar to those observed in \citet{Brasseur2019}.

We investigate the surface conditions during the peak emission times for the largest $\geq$10$^{34}$ flares in our sample. For each flare, we measure the energy emitted in the TESS bandpass during just the 20 second exposure at the flare peak time and scale the result to the HZ distance using \citet{Kopparapu2013} and the stellar mass. We then assume a 9000 K blackbody \citep{Osten2015} to estimate a lower limit on the UVC fluence reaching the surface of an unmagnetized planet in the absence of a significant atmosphere. We note this is an exploratory case designed to probe the relative effects of the peak flux and that the true UVC could be much lower given hazes or an Archean Earth atmosphere \citep{Rugheimer2015, OMalley2019}. We define the UVC to fall within the range 100-280 nm, although we do not consider UVC radiation below 240 nm due to the effectiveness of atmospheric volatiles in blocking these wavelengths \citep{Tilley2019}. We define the FUV to fall within the range 90-170 nm \citep{Loyd2018}.

\begin{figure*}
	\centering
	{
		\includegraphics[trim= 0 0 0 0, width=0.99\textwidth]{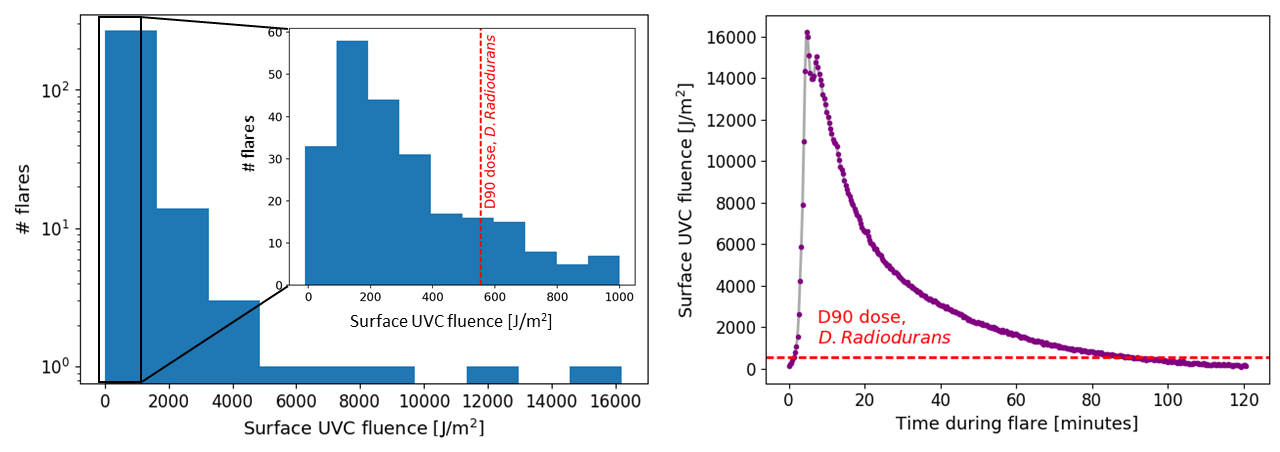}
	}
	\caption{Left: The distribution of instantaneous UVC surface fluences of any HZ planets. Fluences are taken from the 20 second exposure during each flare peak, not from integrating over the duration of the flares. An inset panel shows the region near the 553 J m$^{-2}$ D90 dose for the radiation-hardy \textit{D. Radiodurans}. Note the inset panel has a linear y-scale while the larger panel has a log y scale. UVC surface fluences assume a HZ distance and an unmagnetized planet with no significant UVC absorbers longwards of the $\sim$240 nm haze limit (e.g. \citet{Tilley2019}). Right: The flare with the highest instantaneous surface fluence, reaching 16,200 J m$^{-2}$. The flare was emitted from a 40 Myr old M4 dwarf in Tuc-Hor, UCAC2 14970156 \citep{Kraus2014, Riaz2006}. Binarity has been ruled out for the host with Lucky Imaging \citep{Bergfors2010}. While the peak flux does not reach the 39,000 J m$^{-2}$ needed to inactivate 100\% of \textit{D. Radiodurans}, the total integrated fluence over the duration of the flare does reach this limit.}
	\label{fig:UVC_fluences}
\end{figure*}

We find that peak-time fluences in our sample are distributed with a mean and 1$\sigma$ range of 730$\substack{+820 \\ -210}$ J m$^{-2}$. The largest peak-time fluence reaches 16,200 J m$^{-2}$ (Figure \ref{fig:UVC_fluences}). For reference, 553 J m$^{-2}$ is the D90 dose rate required to inactivate 90\% of a population of the radiation-hardy bacterium \textit{D. Radiodurans} \citep{Gascon1995}. Fully 1/3 of our 10$^{34}$ erg flares reach this limit during the 20 second peak epoch. None of our flare peaks reach the much larger 39,000 J m$^{-2}$ needed to inactivate $\sim$100\% of the population \citep{Abrevaya2020}. Survival curves for the common bacterium \textit{P. Aeruginosa} also obtained from laboratory experiments by \citet{Abrevaya2020} enable us to estimate the survival fractions of this microorganism during our 20 second duration peak fluences. Our estimated fluences would result in survival fractions of $\sim$10$^{-2.1}$ with the largest 20 second peaks resulting in survival fractions of 10$^{-4.9}$. Further work is needed to assess the effects of longer-duration fluences, repeated flaring, and natural selection on population regrowth times \citep{Estrela2020}. We also estimate the top-of-atmosphere far UV (FUV) flux in the HZ by linearly scaling from a large flare observed simultaneously by TESS and the Hubble Space Telescope which emitted 10$^{30.3}$ erg in the FUV and 10$^{30.5}$ in the TESS bandpass \citep{MacGregor2021}. We estimate that 20 second cadence peak-time FUV fluences in our sample are distributed with a mean and 1$\sigma$ range of 2200$\substack{+2400 \\ -600}$ J m$^{-2}$.

\section{Conclusions}\label{conclude}

We have conducted the largest statistical exploration to date of the detailed time-resolved properties of M-dwarf flares in the optical using the new TESS 20 second cadence mode. While previous studies have carefully characterized dozens of large optical flares from a handful of M-dwarf hosts with high time-resolution spectra and photometry (e.g. \citealt{Kowalski2010, Kowalski2013, Kowalski2016, Kowalski2019}) or accurately measured the energy and amplitudes of large numbers of M-dwarf flares in high cadence photometry (e.g. \citet{Jackman2021, Webb2021, Gilbert2021}), we place new constraints on the occurrence rates and detailed properties of substructure within large flares. The vast majority of these events are superflares, or events of $\geq$10$^{33}$ erg. High-cadence observations of large numbers of M-dwarf superflares have been difficult to obtain in the past due to the rarity and stochastic nature of these events. 

The high rate of complex emission in the rise phases of large flares and the range of peak morphologies revealed by 20 second cadence TESS observations need to be reproduced in models of electron acceleration and heating in flares (e.g. \citealt{Allred2006, allred2015, Kowalski2019}). The high occurrence rate of short-period QPPs in our sample also serve to constrain the structure and particle acceleration in flares (e.g. \citealt{Doorsselaere2016, Kupriyanova2020, Zimovets2021} and references therein). Future constraints on the electron velocities from QPPs in TESS flares may also be joined with constraints derived from the millimeter using the flare's spectral index, field strength, and source volume to better understand particle acceleration in flares \citep{MacGregor2021, Ramsay2021}. An increased understanding of flare emission and particle acceleration will in turn propagate to increased accuracy in modeling planetary habitability \citep{Loyd2018, Tilley2019, Chen2021}.

A subset of the superflares reported in this paper were scheduled to be observed simultaneously at 20 second cadence by TESS and at 2 minute cadence by the Evryscope South array of small telescopes \citep{Law2016, Ratzloff2019}. Due to the ongoing COVID-19 pandemic, Cerro Tololo Inter-American Observatory (CTIO) and Evryscope-South were shut down for several months during the TESS observations of the scheduled fields. The Cycle 3 targets that were observed simultaneously with TESS after CTIO reopened and the Cycle 4 targets currently being observed  will be analyzed in future work. We will compare the impulsive properties of the flares at 20 second cadence with the color-temperature evolution at 2 minute cadence to determine if more impulsive peaks correlate with higher color-temperatures.

We also urge further work be done to determine if changes in the time-resolved emission of superflares would lead to measurable short-term changes to planetary atmospheres or bio-signatures. For example, photoprotective biofluorescence has been proposed as a temporal bio-signature caused by large M-dwarf flares \citep{OMalley2019_bio}. The temporal evolution of one superflare from AD Leo \citep{Hawley1991} was employed by \citet{OMalley2019_bio} to model a $\sim$100$\times$ increase to the planet-star contrast at biofluorescent wavelengths. However, it is unclear if or how strongly temporal bio-signatures might differ between superflares with high versus low impulsivity. Based on the wide range of flare morphologies we discover, we hypothesize that differently-shaped flares could lead to different model outcomes.

\section*{Acknowledgements}\label{acknowledge}
We would like to thank the anonymous referee for improving the work. WH would also like to thank Nicholas Law and Cole Tamburri for helpful conversations on flare substructure.
\par WH acknowledges funding support by the TESS Guest Investigator Program GO 3174.
\par This paper includes data collected by the TESS mission. Funding for the TESS mission is provided by the NASA Explorer Program. This work has made use of data from the European Space Agency (ESA) mission {\it Gaia} (\url{https://www.cosmos.esa.int/gaia}), processed by the {\it Gaia} Data Processing and Analysis Consortium (DPAC, \url{https://www.cosmos.esa.int/web/gaia/dpac/consortium}). Funding for the DPAC has been provided by national institutions, in particular the institutions participating in the {\it Gaia} Multilateral Agreement. This research made use of Astropy,\footnote{http://www.astropy.org} a community-developed core Python package for Astronomy \citep{astropy:2013, astropy:2018}, and the NumPy, SciPy, and Matplotlib Python modules \citep{numpyscipy,2020SciPy-NMeth,matplotlib}.

{\it Facilities:} \facility{TESS}

\bibliographystyle{apj}
\bibliography{paper_references}

\vspace{-0.5cm}
\appendix
\vspace{-0.1cm}
\section{A. Unconfirmed QPP Candidates Flagged During Initial Visual Inspection}\label{lowquality_QPPs}
\begin{figure}[H]
    \vspace{-0.4cm}
	\centering
	{
		\includegraphics[trim= 0 20 0 0, width=0.9\textwidth]{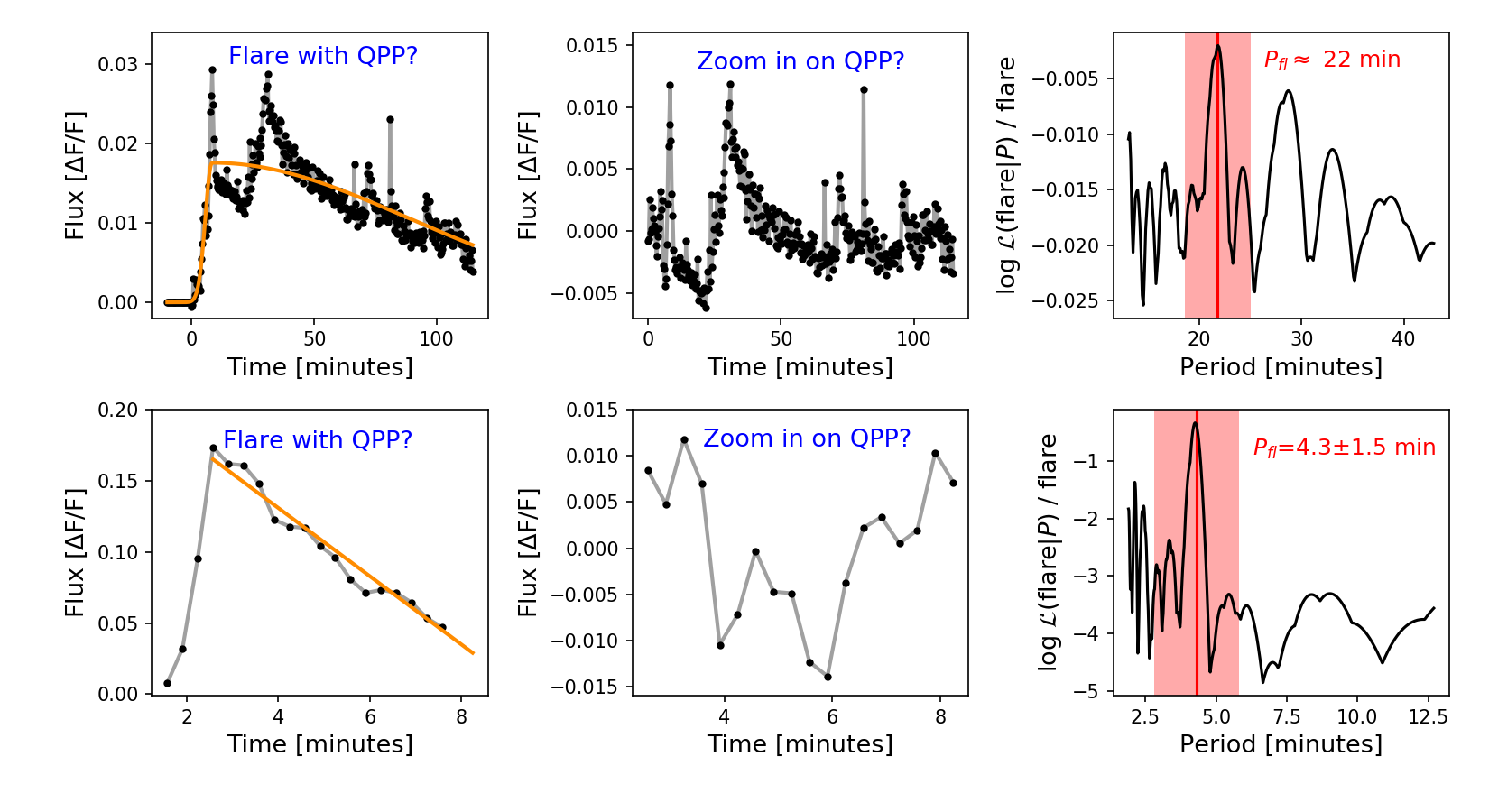}
	}
	\vspace{-0.23cm}
	\caption{Typical examples of the weakest candidate QPPs identified during visual inspection that did not meet the formal 3$\sigma$ cutoff. Some of the lower significance events are very likely to be true QPPs. Others are possibly QPPs but may be random superpositions of peaks in the flare substructure. The top flare shows periodic substructure but is missing an event between the largest peak and the next component that would make the periodicity clearer. The bottom panel candidate has a signal with a low amplitude of variability. Panels are the same as described in Figure \ref{fig:qpp_methods}, except that the Gaussian model in the bottom left panel has been replaced with a linear model.}
	\label{fig:candidate_qpp_examples}
\end{figure}

\vspace{-0.1cm}
In total, 32 candidate QPPs were flagged during visual inspection that did not meet the formal 3$\sigma$ detection threshold outlined in Section \ref{qpp_flares}. Many of these lower significance candidates are true QPPs (e.g. see Figure \ref{fig:multimodal}), while others are probably random superpositions of peaks in the flare substructure. In Figure \ref{fig:candidate_qpp_examples}, we illustrate two of the weakest unconfirmed candidates to aid the reader in understanding the full range of signals flagged during the initial visual inspection. We do not claim these two examples to be true QPPs.

\end{document}